\documentclass[prb,twocolumn,showpacs,showkeys]{revtex4}

\usepackage{graphicx}
\usepackage{subfigure}
\usepackage{amssymb}
\usepackage[dvips]{color}

\begin{document}
\title{Semiconducting (Half-Metallic) Ferromagnetism in Mn(Fe) Substituted 
       Pt and Pd Nitrides}

\author{Abdesalem Houari}
\email[Corresponding authors: \\] {habdslam@yahoo.fr}
\affiliation{Laboratoire de Physique Th\'eorique, 
             D\'epartement de Physique, 
             Universit\'e de B\'ejaia, 
             B\'ejaia, Alg\'erie}

\author{Samir F.~Matar}
\affiliation{CNRS, Universit\'e de Bordeaux, ICMCB, 
             87 avenue du Docteur Albert Schweitzer, 
             33600 Pessac, France}

\author{Volker Eyert}
\email[] {volker@eyert.de} 
\affiliation{Center for Electronic Correlations and Magnetism,
             Institut f\"ur Physik, Universit\"at Augsburg,
             86135 Augsburg, Germany}
\date{\today}
%\maketitle

\begin{abstract}
Using first principles calculations as based on density functional theory, 
we propose a class of so far unexplored diluted ferromagnetic 
semiconductors and half-metals. 
Here, we study the electronic properties of recently synthesized 
$ 4d $ and $ 5d $ transition metal dinitrides. In particular, we
address Mn- and Fe-substitution in PtN$_2$ and PdN$_2$. 
Structural relaxation shows that the resulting ordered compounds, 
Pt$_{0.75}$(Mn,Fe)$_{0.25}$N$_2$ and Pd$_{0.75}$(Mn,Fe)$_{0.25}$N$_2$, 
maintain the cubic crystal symmetry of the parent compounds. 
On substitution, all compounds exhibit long-range ferromagnetic order. 
While both Pt$_{0.75}$Mn$_{0.25}$N$_2$ and Pd$_{0.75}$Mn$_{0.25}$N$_2$ 
are semiconducting, Fe-substitution causes half-metallic behavior for 
both parent materials. 
\end{abstract}
%\pacs{72.25.Ba}% PACS, the Physics and Astronomy
\pacs{71.15.Mb, 71.15.Nc, 71.20-b, 75.10.Lp, 74.25.Ha, 73.43.Cd}
\maketitle

Being known since the beginning of the 20th century, transition metal 
nitrides are considered as an exciting class of materials due to a 
wide range of technological applications. Traditionally, the great 
advantages of these compounds concern their hardness and refractory 
nature \cite{Oyama,Pierson}. However, much attention is currently 
directed towards their electronic, magnetic, and optical properties, 
where fascinating applications are expected.    
Many efforts, experimental as well as theoretical, have been 
made to study the transition metal nitrides \cite{Houari,Houari1,Jhi,Eriksson}.
Until recently none of the noble metal nitrides or the 
platinum group nitrides were known. The first synthesis of platinum 
nitride (Pt-N), under extreme conditions of pressure and temperature, 
was reported only few years ago \cite{Gregor,Crow}. Lateron, several 
other nitrides of different elements (Ir, Os, Ru and Pd) were also 
obtained \cite{Crow1,Young,Yu}. 

There was a debate about the crystal structure and the stoichiometry
of platinum nitride. While a zinc-blende structure was first proposed, 
Crowhurst {\it et al.} demonstrated that this nitride crystallizes 
neither in zinc-blende (PtN: mononitride) nor in fluorite 
(PtN$_2$: dinitride) type structures, which are highly unstable at the 
synthesis conditions ($ P=50 $\,GPa and $ T=2000 $\,K) \cite{Crow}. 
Instead, the authors revealed that the coumpound is a dinitride, hence PtN$_2$, 
and the ground state structure is a cubic pyrite structure. Lateron, these  
authors succeeded in synthesizing IrN$_2$ and PdN$_2$, where the first one 
is found to be in the monoclinic baddeleyite structure \cite{Crow1}. 
Yet, PdN$_2$, which could by synthesized at high pressures but was not 
stable at ambient conditions,  
%As reported in the synthesis experiments \cite{Crow1}, PdN$_2$ is 
%obtained at 58\,GPa. On going down to ambient pressure, the compound 
%is destroyed (most probabely decomposed) at 13 GPa and no stable 
%compound is thus obtained at pressure below. So, at zero pressure,  
%there is no experimantal PdN$_2$ in pyrite structure.)
was proposed to also crystallize in the pyrite structure.  
In a recent theoretical investigation it 
was shown that tetragonal distortions may stabilize PdN$_2$ at ambient 
pressure \cite{Aberg}. Other nitrides (OsN$_2$, RuN$_2$ and RhN$_2$)
have been also obtained and are shown to crystallize in marcasite type
structure \cite{Young,Yu}.

Platinum dinitride has been predicted to have excellent mechanical properties.
The calculated hardness (bulk modulus, shear modulus and other elastic
constants) shows that it is harder than many known hard materials like TiN and
SiC \cite{Yu1,Gou}.  The electronic properties of PtN$_2$ are also
very interesting. Contrary to other transition metal nitrides, which are
almost all metallic, PtN$_2$ is found to be semiconducting, and this could 
make it an important material for optoelectronic applications. 
Band structure calculations as based on density functional theory and 
the local density approximation (LDA) show an indirect band gap of 
$\sim$ 1.5 eV, which is probably somewhat smaller than the experimental 
value due to the tendency of the LDA to underestimate the band gap 
\cite{Yu1,Gou,Young1}. 

In general, pyrite-type compounds have attracted attention since long. 
The dinitrides $ {\rm AN_2} $ (A=C,Si,Ge) were devised in assumed pyrite-type 
structures leading to compounds with peculiar properties, such as the
extreme hardness obtained for $ {\rm CN_2} $ with a bulk modulus of 
405\,GPa. For 
these systems characterized as wide band gap semiconductors, strong 
hybridization of the N $ 2p $ states with the A $ p $ states results 
in a depression of the optical band gap along the C, Si, Ge series 
\cite{cn2,cn2b}. 
Pyrite-type disulfides have also been of considerable interest for 
different reasons \cite{goodenough71a}. 
Semiconducting $ {\rm FeS_2} $ has found widespread attention for 
its application in photovoltaic energy conversion \cite{fes2}. 
$ {\rm ZnS_2} $ is a diamagnetic insulator. Substitution of Zn for 
Fe in iron pyrite has thus been used to tune the optical band gap 
in order to enhance the response to the solar spectrum \cite{fes2zn}. 
While $ {\rm FeS_2} $ is a van Vleck paramagnet, 
metallic $ {\rm CoS_2} $ displays long-range ferromagnetic order. 
In contrast, $ {\rm NiS_2} $ is an antiferromagnetic insulator, where 
the insulating behaviour has been attributed to the presence of strong 
electronic correlations. 

Our present work is focused especially on the electronic and magnetic 
properties of substituted PtN$_2$ and PdN$_2$. We demonstrate that 
substitution of the non-magnetic $ 4d $- and $ 5d $-transition 
metal ions by the magnetic $ 3d $-ions Mn and Fe may lead to 
semiconducting and half-metallic ferromagnetism, respectively. 

In our investigation, we first consider Mn-substitution in PtN$_2$ and 
PdN$_2$. For the latter compound, we assumed a cubic pyrite crystal 
structure  as for PtN$_2$. This assumption is based on the fact that 
experimentally PdN$_2$ is actually synthesized in cubic pyrite structure 
at high pressure conditions, even though it is not stable at 
ambient pressure \cite{Crow1}. Replacing one of the four Pt and Pd 
by magnetic Mn leads to Pt$_{0.75}$Mn$_{0.25}$N$_2$ and 
Pd$_{0.75}$Mn$_{0.25}$N$_2$, respectively. To check if the cubic 
symmetry is maintained on Mn-substitution, a quantum molecular dynamics  
relaxation has been performed using the Siesta {\it ab initio} simulation 
package with norm-conserving pseudopotentials \cite{Siesta,PP}. 
%In doing so, we employed the LDA, which was found to give good agreement 
% with experimental TN$_2$ (T = Pt, Ir, Os and Pd) lattice constants.   
Both atomic positions and 
cell shape were included in the relaxation process. As a result, neither 
Pt$_{0.75}$Mn$_{0.25}$N$_2$ nor Pd$_{0.75}$Mn$_{0.25}$N$_2$ display 
any deviations from cubic symmetry and the atoms remain nearly at the 
positions of the pure compound. In particular, the internal nitrogen 
parameters are almost unchanged after Mn-substitution in both $ {\rm PtN_2} $ 
and $ {\rm PdN_2} $. The changes in the nitrogen positions are within 
0.07\,\AA. To be specific, in $ {\rm PtN_2} $ the internal nitrogen 
parameter changes from 0.415 (as given in Ref.\ \onlinecite {Crow}) to 0.416 
after the relaxation of the substituted system.

In a second step, full potential augmented spherical wave (FPASW) 
calculations were carried out \cite{aswrev,aswbook} in order to address the 
electronic properties of all compounds under study. To start with, 
we recalculated the equilibrium lattice constant. From non-spin polarized 
LDA calculations, we obtained a lattice parameter of 
$ a_{NM} = 4.79 $\,\AA \ for $ {\rm Pt_{0.75}Mn_{0.25}N_2} $. Taking into 
account spin polarization led to a slightly larger value of 
$ a_{FM} = 4.82 $\,\AA, 
with the ferromagnetic state being more stable than the non-magnetic one. 
It is important to note that the values obtained for the lattice constant 
of Pt$_{0.75}$Mn$_{0.25}$N$_2$ resemble that of PtN$_2$, which is 
$ a = 4.80 $\,\AA, and confirm the molecular dynamics result. To conclude  
from both sets of calculations, not only the cubic symmetry is preserved 
after Mn-substitution, but even the lattice constant is almost not
affected. The negligible changes of the structure can be understood 
from the fact that only one out of four Pt atoms is replaced and thus 
the platinum network is affected by the substitution only to a small 
degree. Motivated by these findings, we decided to perform the 
calculations for Pd$_{0.75}$Mn$_{0.25}$N$_2$ using the same lattice 
constant as for PdN$_2$, i.e.\ $ a = 4.75 $\,\AA \ (see also the 
discussion below). 

\begin{figure}[htbp]
%\subfigure{\includegraphics[width=0.45\textwidth]{Fig1a}} 
%\subfigure{\includegraphics[width=0.95\textwidth]{Fig1b}}
\includegraphics[width=\columnwidth]{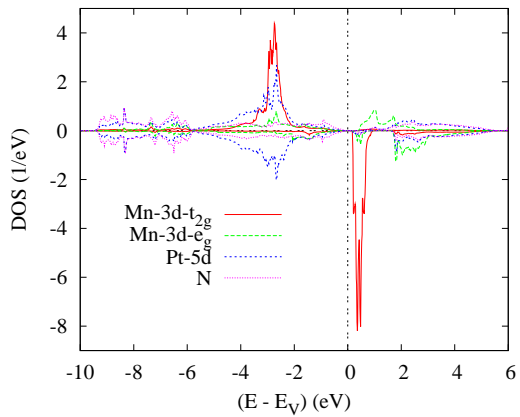}
\caption{(Color online) 
         Partial DOS of Pt$_{0.75}$Mn$_{0.25}$N$_2$. Here and in all 
         subsequent figures, $ t_{2g} $ and $ e_g $ orbitals are referred 
         to a rotated coordinate system with the Cartesian axes pointing 
         along the metal-nitrogen bonds.}
\label{fig:dosptmn}
\end{figure}

\begin{figure}[htbp]
%\subfigure{\includegraphics[width=0.45\textwidth]{Fig2a}} 
%\subfigure{\includegraphics[width=0.95\textwidth]{Fig2b}}
\includegraphics[width=\columnwidth]{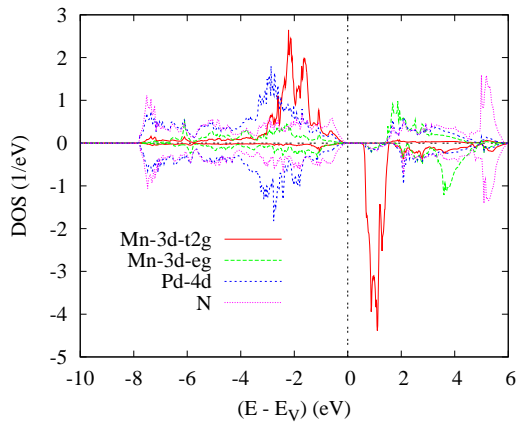}
\caption{(Color online) 
         Partial DOS of Pd$_{0.75}$Mn$_{0.25}$N$_2$.}
\label{fig:dospdmn}
\end{figure}

Subsequently, the electronic structures of the Mn-substituted compounds  
were analyzed in terms of the projected densities of states as arising 
from the FPASW calculations. However, for the Mn-substituted systems the 
LDA results bear some ambiguity. To be specific, we obtain semiconducting 
behavior for Pd$_{0.75}$Mn$_{0.25}$N$_2$, whereas 
Pt$_{0.75}$Mn$_{0.25}$N$_2$ is at the verge of being a semiconductor but 
displays a small band overlap. In order to check these findings, we 
additionally performed a set of calculations based on the GGA \cite{GGA}. 
They resulted in semiconducting ferromagnetic ground states for both 
compounds with indirect band gaps of 0.17\,eV and 0.42\,eV for 
Pt$_{0.75}$Mn$_{0.25}$N$_2$ and Pd$_{0.75}$Mn$_{0.25}$N$_2$, respectively. 
The corresponding partial densities of states (DOS) are illustrated 
in Figs.\ \ref{fig:dosptmn}
and \ref{fig:dospdmn}. 
The spectrum falls essentially in four parts. While the low-energy range 
from $ -9 $ to $ -6 $\,eV and from $ -8 $ to $ -4 $\,eV for 
$ {\rm Pt_{0.75}Mn_{0.25}N_2} $ and $ {\rm Pd_{0.75}Mn_{0.25}N_2} $, 
respectively, is dominated by the N $ 2p $ states, the upper 
valence band is formed mainly by the $ t_{2g} $ manifolds of the transition 
metal $ d $ states. 
%In this context we point out that the octahedral 
%environment of the transition metal atoms is slightly distorted. For 
%this reason, the crystal field splitting into states 
%of $ t_{2g} $ and $ e_g $ symmetry is not perfect and small admixtures 
%of the respective other manifoled are observed. 
%\textcolor{red}{Abdeslam, in generating the figures, did you account for 
%the rotation of the octahedra?} 
%\textcolor{blue}{Yes, indeed, I followed the same procedure of plotting 
%like FeS2 in the userguide, by taking R(23)(1,1,0)*R(17)(0,0,1) as 
%rotation symbol}
For the spin-minority bands, the situation is slightly more complicated. 
Whereas the Pd $ 4d $ and Pt $ 5d $ states are fully occupied 
and found in the same energy range as their spin-majority counterparts, 
the Mn $ 3d $ $ t_{2g} $ states experience strong exchange splitting. 
As a result, the Mn $ 3d $ $ t_{2g} $ spin-down states form the lower 
conduction band of this spin channel and a magnetic moment of 3$\mu_B$ 
is found at these atoms. In contrast, spin polarizations of Pd, Pt, and 
N are negligible. Finally, the remaining conduction band states can be 
attributed to the transition metal $ d $ states of $ e_g $ symmetry. 
Since the latter form $ \sigma $-type bonds with the N $ 2p $ states, 
we find a considerable admixture of both types of states in the lower 
valence and upper conduction band. This admixture is much smaller for 
the bands between $ -6 $ and $ +1 $\,eV and $ -4 $ to $ +2 $\,eV, 
respectively, which are of $ t_{2g} $ symmetry 
and form less strong $ \pi $ bonds. In passing we mention the albeit 
small band gaps, which make both Mn-substituted compounds semiconducting. 
Yet, we note that LDA and GGA underestimate the optical band gap, which 
might thus be considerably larger in reality. In order to check this,  
we performed additional GGA+$ U $ calculations for 
$ {\rm Pt_{0.75}Mn_{0.25}N_2} $. While there were no qualitative 
changes, both the exchange splitting of the Mn $ 3d $ $ t_{2g} $ states 
and the optical band gap increased considerably.

The second substitution that we considered was the replacement of 
Pt and Pd by iron, leading 
to the ordered compounds Pt$_{0.75}$Fe$_{0.25}$N$_2$ and 
Pd$_{0.75}$Fe$_{0.25}$N$_2$. Our procedure was the same as for 
Mn-substitution. Molecular dynamics relaxations using the Siesta code
(with the same calculations details cited above)w
were performed including relaxation of both the atomic positions and 
the cell shape. As in the Mn-case we found that the cubic symmetry is 
not broken on Fe-substitution and that even the internal nitrogen 
parameter remained essentially unchanged. 

For the FPASW calculations performed in a second step in order to 
address the electronic and magnetic properties, we followed the 
procedure already adopted for $ {\rm PdN_2} $ and used the lattice 
constants of the pure systems also for the substituted materials. 
In this case, our procedure was justified by an additional 
recalculation of the equilibrium lattice constant for 
$ {\rm Pd_{0.75}Fe_{0.25}N_2} $. As a result, values of 
$ a_{NM} = 4.742 $\,\AA \ and $ a_{FM} = 4.749 $\,\AA \ were 
obtained as arising from non-spin polarized and spin polarized 
calculations, respectively. The latter value is almost identical 
to the value of $ a = 4.75 $\,\AA \ of pure $ {\rm PdN_2} $. 

Again, the LDA results bear some ambiguity as they led to half-metallic 
behavior for Pd$_{0.75}$Fe$_{0.25}$N$_2$ but metallicity of both spin 
channels for Pt$_{0.75}$Fe$_{0.25}$N$_2$. Yet, the spin-majority density 
of states 
at the Fermi energy was found to be very small. The problem could 
again be resolved by GGA calculations, which render both substituted 
materials half-metallic. Both compounds exhibit stable magnetic order 
with magnetic moments of 2.0\,$\mu_B$ located almost completely at  
the iron atoms. 

\begin{figure}[htbp]
%\subfigure{\includegraphics[width=0.45\textwidth]{Fig3a}} 
%\subfigure{\includegraphics[width=0.95\textwidth]{Fig3b}}
\includegraphics[width=\columnwidth]{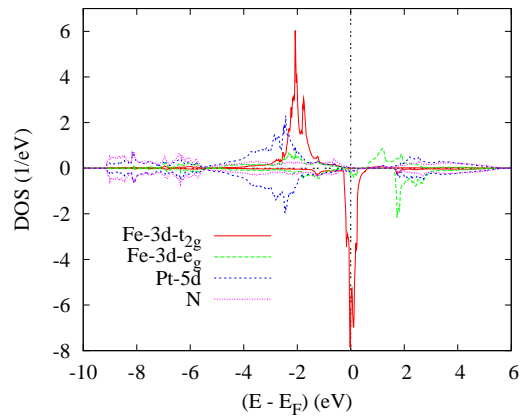}
\caption{(Color online) 
         Partial DOS of Pt$_{0.75}$Fe$_{0.25}$N$_2$.}
\label{fig:dosptfe}
\end{figure}

\begin{figure}[htbp]
%\subfigure{\includegraphics[width=0.45\textwidth]{Fig4a}} 
%\subfigure{\includegraphics[width=0.95\textwidth]{Fig4b}}
\includegraphics[width=\columnwidth]{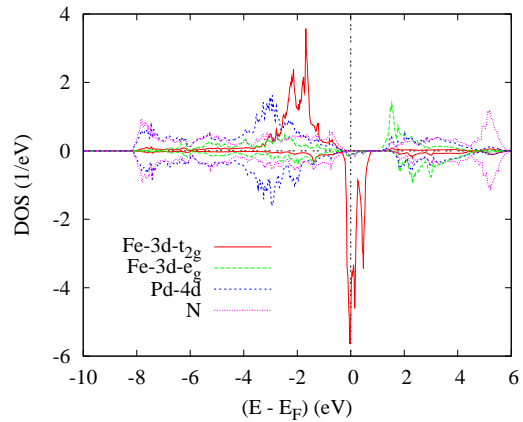}
\caption{(Color online) 
         Partial DOS of Pd$_{0.75}$Fe$_{0.25}$N$_2$.}
\label{fig:dospdfe}
\end{figure}

The electronic structure and partial DOS of the two compounds as 
arising from the spin-polarized ferromagnetic calculations are 
illustrated in Figs.\ \ref{fig:dosptfe}
and \ref{fig:dospdfe}. 
The gross features of the partial densities of states are the same 
as for the Mn-substituted compounds. Differences are due to the 
smaller magnetic moments of the Fe atoms, which lead to reduced 
exchange splittings of the $ 3d $ $ t_{2g} $ states. As a consequence, 
the respective spin-majority bands are shifted to higher energies 
as compared to the $ d $ states of the Pt and Pd matrix. In addition, 
the Fe spin-minority bands are shifted to lower energies as compared 
to the Mn-systems due to the increased electron count. As a result, 
the semiconducting behavior is lost and a metallic spin-down channel 
found. Again, these results were qualitatively confirmed by additional 
GGA+$ U $ calculations, which revealed an increase of the exchange 
splitting of the Fe $ 3d $ $ t_{2g} $ states as well as the band 
gap of the spin-majority channel.
  
In passing, we mention additional spin polarized calculations, which
were performed Pt$_{0.75}$Mn$_{0.25}$N$_2$ in order to check for long
range antiferromagnetic order. For these calculations, we used a
tetragonal structure arising from doubling the cubic cell along
the $c$ axis. As a result, an antiferromagnetic and again 
semiconducting solution was found albeit with a total energy, which
by about 7 mRy/f.u. higher than that of the ferromagnetic ground
state.    

In summary, based on our first principles investigation we propose the 
existence of so far unexplored diluted ferromagnetic semiconductors 
and half-metals. 
These materials arise from substituting magnetic $ 3d $ ions (Mn, Fe) 
in the non-magnetic dinitrides $ {\rm PtN_2} $ and $ {\rm PdN_2} $. 
According to molecular dynamics calculations, the ordered compounds 
A$_{0.75}$B$_{0.25}$N$_2$, where $ {\rm A = Pt, Pd} $ and 
$ {\rm B = Mn, Fe} $, preserve the cubic pyrite structure of their 
parent compounds. On substitution, stable long-range ferromagnetic 
order results with magnetic moments of $ 3 \mu_B $ and $ 2 \mu_B $, 
which are well localized at the Mn- and Fe-sites, respectievly. 
While Mn-substitution leads to semiconducting behavior, introduction of 
iron causes the substituted compounds to be half-metallic. Our results 
still await experimental confirmation.

\begin{acknowledgements}
This work was supported by the Deutsche Forschungsgemeinschaft through TRR 80.
\end{acknowledgements}

{}

%%%%%%%%%%%%%%%%%%%%%%%%%%%%%%%%%%%%%%%%%%%%%%%%%%%%%%%%%%%%%%%%%%%%%%%%

\end{document}